\documentclass[iop]{emulateapj}

\usepackage{graphicx}
\usepackage{epstopdf}
\usepackage{hyperref}
\usepackage{url}
\usepackage{breakurl}
\usepackage{amsmath}
\usepackage{rotating}
\usepackage{longtable}
\usepackage{multirow}
\usepackage{ctable}

\expandafter\def\expandafter\UrlBreaks\expandafter{\UrlBreaks
  \do\a\do\b\do\c\do\d\do\e\do\f\do\g\do\h\do\i\do\j%
  \do\k\do\l\do\m\do\n\do\o\do\p\do\q\do\r\do\s\do\t%
  \do\u\do\v\do\w\do\x\do\y\do\z\do\A\do\B\do\C\do\D%
  \do\E\do\F\do\G\do\H\do\I\do\J\do\K\do\L\do\M\do\N%
  \do\O\do\P\do\Q\do\R\do\S\do\T\do\U\do\V\do\W\do\X%
  \do\Y\do\Z}

\newcommand{\kepler}{\emph{Kepler}}

\slugcomment{Submitted to the Astrophysical Journal, Accepted May 2nd 2016}

\shorttitle{Measuring the Kepler Pipeline Detection Efficiency}
\shortauthors{Christiansen et al.}

\begin{document}


\title{Measuring Transit Signal Recovery in the Kepler Pipeline. III. Completeness of the Q1--Q17 DR24 Planet Candidate Catalogue, with Important Caveats for Occurrence Rate Calculations}

\author{Jessie L. Christiansen$^1$}
\author{Bruce D. Clarke$^2$}
\author{Christopher J. Burke$^2$}
\author{Jon M. Jenkins$^3$}
\author{Stephen T. Bryson$^3$}
\author{Jeffrey L. Coughlin$^2$}
\author{Fergal Mullally$^2$}
\author{Susan E. Thompson$^2$}
\author{Joseph D. Twicken$^2$}
\author{Natalie M. Batalha$^3$}
\author{Michael R. Haas$^3$}
\author{Joseph Catanzarite$^2$}
\author{Jennifer R. Campbell$^4$}
\author{AKM Kamal Uddin$^4$}
\author{Khadeejah Zamudio$^4$}
\author{Jeffrey C. Smith$^2$}
\author{Christopher E. Henze$^3$}
\email{jessie.christiansen@caltech.edu}
\affil{$^1$NASA Exoplanet Science Institute, California Institute of Technology, M/S 100-22, 770 S. Wilson Ave, Pasadena, CA 91106, USA}
\affil{$^2$SETI Institute/NASA Ames Research Center, Moffett Field, CA 94035, USA}
\affil{$^3$NASA Ames Research Center, Moffett Field, CA 94035, USA}
\affil{$^4$Wyle Laboratories/NASA Ames Research Center, Moffett Field, CA 94035, USA}




\begin{abstract}

With each new version of the \kepler\ pipeline and resulting planet candidate catalogue, an updated measurement of the underlying planet population can only be recovered with an corresponding measurement of the \kepler\ pipeline detection efficiency. Here, we present measurements of the sensitivity of the pipeline (version 9.2) used to generate the Q1--Q17 DR24 planet candidate catalog (Coughlin et al. 2016). We measure this by injecting simulated transiting planets into the pixel-level data of 159,013 targets across the entire \kepler\ focal plane, and examining the recovery rate. Unlike previous versions of the \kepler\ pipeline, we find a strong period dependence in the measured detection efficiency, with longer ($>$40 day) periods having a significantly lower detectability than shorter periods, introduced in part by an incorrectly implemented veto. Consequently, the sensitivity of the 9.2 pipeline cannot be cast as a simple one-dimensional function of the signal strength of the candidate planet signal as was possible for previous versions of the pipeline. We report on the implications for occurrence rate calculations based on the Q1--Q17 DR24 planet candidate catalog and offer important caveats and recommendations for performing such calculations. As before, we make available the entire table of injected planet parameters and whether they were recovered by the pipeline, enabling readers to derive the pipeline detection sensitivity in the planet and/or stellar parameter space of their choice. 

\end{abstract}



\section{Introduction}
\label{sec:intro}

The primary goal of the NASA \kepler\ Mission is to measure $\eta_{\Earth}$, the frequency of Earth-size planets in the habitable zone of Sun-like stars. En route to that goal, a larger picture of the underlying planet population has emerged, covering large swathes of planet and stellar host parameter space; recent examples include measurements of the frequency of hot Jupiters \citep[e.g.][]{Santerne2015}, the frequency of Venus-analogues \citep{Kane2014}, and the frequency of Earth-size planets orbiting M-dwarfs \citep{Dressing2015}. 

The most recent advance towards measuring $\eta_{\Earth}$ by the \kepler\ project was presented in \citet{Burke2015}. They use the Q1--Q16 planet catalogue of \citet{Mullally2015}, based on the first 47 months of \kepler\ data, and examine the occurrence rate of planets with radii 0.75--2.5$R_{\oplus}$ and orbital periods 50--300 days around GK dwarf stars. An important improvement in their calculation was the inclusion of the first direct measurement of the detection efficiency of the \kepler\ pipeline used to generate the planet candidate catalogue, presented in \citet{Christiansen2015a}. Understanding the magnitude of the false negative rate, i.e. how many planets were missed in the analysis that would otherwise be expected to be detected, is an essential ingredient in robust occurrence rate calculations. In fact, \citet{Burke2015} demonstrate that changing the assumption of the false negative rate is one of the largest sources of systematic uncertainties in the final occurrence rate error budget. It is also necessary to estimate the false positive rate of the planet candidate catalogue, i.e. the rate at which the candidate population is polluted by other signals such as eclipsing binaries \citep{Bryson2013,Coughlin2014}, variable stars \citep{Thompson2015} and instrumental artifacts \citep{Mullally2016}. We do not examine the false positive rate in this study.

With each refinement of the \kepler\ pipeline and each subsequently regenerated planet candidate catalogue, our assumptions about the detection efficiency must be revisited. Here we continue our efforts to empirically characterise the sensitivity of the \kepler\ pipeline by performing large-scale injections of simulated transiting planet signals and examining the recovery statistics. In our previous transit injection experiments, we first tested one quarter (Q3, 89 days) of data across all 84 CCD channels, in order to examine whether the initial aperture photometry and subsequent co-trending processes in the pipeline systematically altered individual transit events in any way \citep{Christiansen2013}. The conclusion was that, for transits not falling within two days of a long data gap, the pipeline preserved the depth of injected transit signals at the 99.7\% level; i.e. there was no decrease in the depths or signal strength. See that paper for a description of the pipeline processes which can affect individual transit signals.

In our second experiment, we tested four quarters (Q9--Q12) of data across 15 CCD channels ($\sim$10,000 targets), to examine the recovery rate of transit signal trains with periods up to 180 days. The simulated transit signals were processed through almost the complete \kepler\ pipeline, and as closely as possible the pipeline version matched that used to generate the Q1--Q16 catalogue of Kepler Objects of Interest \citep{Mullally2015}. There we concluded that the detection efficiency of the pipeline could be described as a function of the strength of the signal train by a $\Gamma$ cumulative distribution function \citep{Christiansen2015a}, although the fit coefficients varied broadly as a function of stellar type. This measurement of the detection efficiency was then used by \citet{Burke2015} in their occurrence rate calculation described above.

Here we describe the third transit injection experiment, which tests the entire \kepler\ observing baseline (Q1--Q17) for the first time, across all 84 CCD channels. It was performed to measure the sensitivity of the \kepler\ pipeline used to generate the Q1--Q17 Data Release 24 (DR24) catalogue of Kepler Objects of Interest \citep{Coughlin2015} available at the NASA Exoplanet Archive \citep{Akeson2013}\footnote{\url{http://exoplanetarchive.ipac.caltech.edu}}. Some preliminary results from this experiment were presented in \citet{Christiansen2015b}; here we expand on that analysis. In Section 2 we outline the changes to the \kepler\ pipeline and the potential impacts on the detection efficiency. In Section \ref{sec:design} we describe the transit injection experiment designed to characterise the impact, and in Section \ref{sec:results} we examine the resulting detection efficiency. We discuss the detection efficiency that was recovered and the implications for occurrence rate calculations performed with the Q1--Q17 DR24 planet candidate catalogue. In particular, the prescription outlined in \citet{Burke2015} and \citet{Christiansen2015a} for the previous version of the pipeline is not immediately applicable to this version.  


\section{Kepler Pipeline - Updates for SOC 9.2}
\label{sec:pipeline} 

The Q1--Q17 DR24 planet candidate catalogue was the first catalogue produced with a single uniform version of the \kepler\ pipeline, i.e. Science Operations Center (SOC) version 9.2. The pipeline has been described in detail in a series of papers; for an overview see \citet{Jenkins2010a} and Figure 1 therein. In summary, there are five modules: 
\begin{enumerate}
\item Calibration (CAL: calibration of raw pixels; Quintana et al. 2010)\nocite{Quintana2010}, 
\item Photometric Analysis (PA: construction of the initial flux time series from the optimal aperture for each target; Twicken et al. 2010)\nocite{Twicken2010}, 
\item Pre-Search Data Conditioning (PDC: removal of common systematic signals from the flux time series; Smith et al. 2012, Stumpe et al. 2012, Stumpe et al. 2014)\nocite{Smith2012,Stumpe2012,Stumpe2014}, 
\item Transiting Planet Search (TPS:  searching the light curves for periodic transit signals; Jenkins 2002, Jenkins et al. 2010b, Seader et al. 2013, Tenenbaum et al. 2013, Tenenbaum et al. 2014, Seader et al. 2015)\nocite{Jenkins2002,Jenkins2010b,Seader2013,Tenenbaum2013,Tenenbaum2014,Seader2015a}, and \
\item Data Validation (DV: examination and validation of the resulting candidate signals against a suite of diagnostic tests; Wu et al. 2010)\nocite{Wu2010}. 
\end{enumerate}

Some of the potential areas for signal loss in the pipeline prior to transit detection are described in \citet{Christiansen2013}. However, after a periodic transit signal has been detected by the pipeline, exhibiting at least three transits, with a measured detection statistic (a measure of the signal strength called the Multiple Event Statistic, MES) above the \kepler\ pipeline threshold of 7.1$\sigma$ (Jenkins 2002), it must pass additional checks. These vetoes are included in the pipeline to reduce the high false alarm rate of `signals' that are above the pipeline threshold but are caused by noisy artefacts in the light curves; in the experiment described below the vetoes reduced the number of light curves generating detections from $\sim$150,000 (out of 198,000 light curves) to $\sim$50,000. However, these vetoes can also remove legitimate transit signals, and part of our aim is to quantify the extent to which this occurs. The vetoes include: (i) examining the consistency between the depths of the individual transit events comprising the signal train to eliminate a false alarm caused by, for instance, folding one deep `transit' onto two shallow deviations in the flux time series \citep{Tenenbaum2013}; (ii) comparing the shape of the folded transit event to modelled transit events (as compared to box-shaped signals), in order to penalise systematic decreases in depth that are not transit-like in nature \citep{Seader2013,Seader2015a}; (iii) and most recently in SOC version 9.2 by the introduction of the statistical bootstrap metric \citep{Seader2015a,Jenkins2015}.

The 7.1$\sigma$ threshold used by the pipeline (hereafter referred to as the pipeline-based detection threshold) was chosen to achieve a false alarm rate of 6.24$\times10^{-13}$ on data which, when whitened, was dominated by Gaussian noise. The statistical bootstrap metric drops the assumption that the data have been perfectly whitened and, for each light curve, analyzes the distribution of the out-of-transit data points to estimate the statistical significance of each candidate signal. It then calculates an updated estimate of the threshold on a target-by-target basis required to achieve the requisite false alarm rate of 6.24$\times10^{-13}$ (hereafter referred to as the bootstrap detection threshold). The goal was to achieve a uniform false alarm rate in the presence of non-Gaussian noise on the observations. While this new metric was effective in reducing the number of false alarms, the implementation contained a flaw that produced incorrect threshold values with a high level of scatter in both the significance and threshold estimates. Rather than achieving a search to a more uniform false alarm rate (the design goal for TPS), this flaw contributed to a period-dependent, non-uniform search with respect to the control of the false alarm rate. One of the goals of this transit injection experiment was to quantify the impact of this new behaviour on the pipeline detection efficiency, discussed further in Section \ref{sec:results}.


\section{Experiment Design}
\label{sec:design}

The average detection efficiency describes the likelihood that the Kepler pipeline would successfully recover a given transit signal. To measure this property, we perform a Monte Carlo experiment where we inject the signatures of simulated transiting planets around 198,154 target stars, one per star, across the focal plane, starting with the Q1--Q17 DR 24 calibrated pixels. The simulated transits are generated using the \citet{Mandel02} model, and have orbital periods ranging uniformly from 0.5 to 500 days and planet radii ranging uniformly from 0.25 to 7.0 Re. Orbital eccentricity is set to 0, and the impact parameter is drawn from a uniform distribution between 0 and 1. We then process the modified pixels through the data reduction and planet search pipeline as usual (modules PA through DV). As in our previous experiments, the only departure from standard operations is that the motion polynomials (used for calculating the location of the target) and the cotrending basis vectors (used in the correction of systematic errors) are generated from a `clean' pipeline run that does not contain injected transit signals. This is to avoid corruption from the presence of the injected transits, since the motion polynomials and cotrending basis vectors are generated from the data themselves, and will be distorted by the addition of simulated transit signals on every target. Of the injections, 159,013 resulted in three or more injected transits (the minimum required for detection by the pipeline) and were used for the subsequent analysis. The full table of injected parameters for all 159,013 injections is hosted at the NASA Exoplanet Archive\footnote{\url{http://exoplanetarchive.ipac.caltech.edu/docs/DR24-Pipeline-Detection-Efficiency-Table.txt}}; a sample is included here in Table \ref{tab:results} for illustration of content.

Of the 159,013 targets, most (129,611 across 68 channels) have the simulated transit signal injected at the nominal\footnote{Tests indicate our injections lie within 0.4 arcseconds (0.1 pixels) of the target pixel response function centre of light $\sim$90\% of the time. The amount of flux removed from the target aperture is calculated after the signal is injected, therefore small stochastic errors in the location of the injected flux will not affect the resulting calculations.} target location on the CCD, thereby mimicking a planet orbiting the specified target.  The remaining targets (29,402 across 16 channels chosen to broadly sample the \kepler focal plane and CCD characteristics) have their simulated signal injected slightly offset (0.4--4 arcseconds, or 0.1--1 \kepler\ pixels) from the target location, thereby mimicking a foreground or background transiting planet or eclipsing binary along the line of sight. The offset limits were chosen based on previous transit injection tests---below 0.4 arc seconds, the ability of the pipeline to accurately measure the location of the photocenter of light is dominated by the uncertainty introduced by averaging locations over multiple quarters (see, e.g. Section 3.4.1 of \citet{Bryson2013}). Above 4 arc seconds, the pipeline can readily identify offsets for transit signals $>3\sigma$ significance. The presence and size of these centroid offsets are indicated in Table \ref{tab:results} by a flag in the OF (offset flag) column, where a value of 1 indicates an offset was injected, and the Offset column, where the offset is given in arc seconds, respectively. These injections can be used to test the ability of the pipeline to discriminate between this type of false positive signal and real planetary signals (Mullally et al. submitted). 

\begin{sidewaystable}[h]
\footnotesize
\centering
\caption{Injected and recovered parameters of the injected transiting planets. The full table (159,013 rows) is available from the NASA Exoplanet Archive. The columns are as follows: (i)~KepID: the Kepler ID of the target; (ii)~SG: the sky group in which the target is located; (iii)~$P$: the orbital period of the injected transit signal in days; (iv)~$T_0$: the epoch of the injected transit signal, given in BMJD; (v)~$T_{\rm d}$: the depth of the injected transit signal in parts per million (ppm); (vi)~$t_{14}$: the duration of the injected transit in hours; (vii)~$b$: the impact parameter of the injected transit signal; (viii)~$r$: the ratio of the planet radius to the stellar radius for the injected signal; (ix)~$k$: the ratio of the semi-major axis of the planetary orbit to the stellar radius for the injected signal; (x)~OF: a flag indicating whether the transit signal was injected on the target star (0) or offset from the target star (1) to mimic a false positive; (xi)~Offset: for targets injected off the target source, the distance from the target source location to the location of the injected signal in arcseconds; (xii)~E\_MES: the expected multiple event statistic (MES) of the injected transit signal; (xiii)~RF: a flag indicating successful (1) or unsuccessful (0) recovery of the injected signal by the pipeline. A value of 2 indicates that the signal was recovered by the pipeline at an integer alias of the injected period. Columns (xiv)--(xxi) are only complete for entries with successful recoveries. Column (xiv)~R\_MES: the maximum MES measured by the pipeline on the recovered signal; (xv)~R($P$): the orbital period of the recovered signal in days; (xvi)~R($T_0$): the epoch of the recovered signal, given in BJMD; (xvii)~R$(T_{\rm d}$): the central transit depth of the recovered signal in parts per million (ppm); (xviii)~R($t_{14}$): the transit duration of the recovered signal in hours; (xix)~R($b$): the impact parameter of the recovered signal; (xx)~R($r$): the ratio of the planet radius to the stellar radius for the recovered signal; and (xxi)~R($k$): the ratio of the semi-major axis of the planetary orbit to the stellar radius for the recovered signal.}
\begin{tabular}{lllllllllllllllllllll}
\hline
\hline
\footnotesize{KepID} & \footnotesize{SG} & \footnotesize{$P$} & \footnotesize{$T_0$} &  \footnotesize{$T_{\rm d}$} & \footnotesize{$t_{14}$} & \footnotesize{$b$} & \footnotesize{$r$} & \footnotesize{$k$} &  \footnotesize{OF} & \footnotesize{Offset} & \footnotesize{E\_MES} & \footnotesize{RF} & \footnotesize{R\_MES} & \footnotesize{R($P$)} & \footnotesize{R($T_0$)} &  \footnotesize{R$(T_{\rm d}$)} & \footnotesize{R($t_{14}$)} & \footnotesize{R($b$)} & \footnotesize{R($r$)} & \footnotesize{R($k$)}\\
\footnotesize                     &                                             & \footnotesize{days}    & \footnotesize{BMJD} & \footnotesize{ppm}                & \footnotesize{hrs}                      &                                   &\                                                  &                                           &                                              &   $^{\prime\prime}$                                    &                                                      &                                                       &                                           &  \footnotesize{days}       &  \footnotesize{BMJD}    &  \footnotesize{ppm}                         & \footnotesize{hrs}                           &                                          &                                                            &                                                \\
\hline
5344302 & 50 &   7.1908 & 54900.0323 &  287 &  3.25 & 0.1965 & 0.0154 &  16.861 & 0 & 0.0000 & 10.2854 & 1 &  9.6179 &   7.1908 & 54964.7572 &  203 & 3.37 & 0.245 & 0.0131 &  16.003 \\
5344312 & 50 & 185.1781 & 54982.5886 &  539 & 10.20 & 0.3731 & 0.0214 & 131.856 & 0 & 0.0000 &  8.2778 & 0 &    null &     null &       null & null & null &  null &   null &    null \\
5344344 & 50 & 154.5847 & 55025.1722 &  817 &  5.79 & 0.7521 & 0.0286 & 143.186 & 0 & 0.0000 & 11.6632 & 1 & 12.2291 & 154.5826 & 55025.1869 &  678 & 5.10 & 0.000 & 0.0237 & 237.011 \\
5344350 & 50 & 323.1424 & 55105.1022 & 3500 &  5.37 & 0.5514 & 0.0552 & 413.223 & 0 & 0.0000 & 17.0070 & 0 &    null &     null &       null & null & null &  null &   null &    null \\
5344409 & 50 & 305.1754 & 55023.4717 &  234 & 10.37 & 0.0246 & 0.0138 & 227.856 & 0 & 0.0000 &  2.4976 & 0 &    null &     null &       null & null & null &  null &   null &    null \\
5344412 & 50 &  26.6892 & 54908.2191 &  290 &  3.28 & 0.6336 & 0.0163 &  49.437 & 0 & 0.0000 &  6.5642 & 0 &    null &     null &       null & null & null &  null &   null &    null \\
5344420 & 50 &  34.1909 & 54905.8020 & 3125 &  4.90 & 0.3519 & 0.0511 &  52.817 & 0 & 0.0000 & 44.9867 & 1 & 39.5125 &  34.1910 & 54974.1793 & 2803 & 4.80 & 0.398 & 0.0487 &  52.794 \\
11956865  &         3   & 109.6892  & 54951.7778     &    1480    &   7.26    &  0.0608    &   0.0341    &   119.171        &               1            & 1.9081      &      9.8913         &     0      &        null      &    null      &      null     &    null     &   null   &      null      &   null     &     null\\
   11956938   &        3   & 402.3242   &  55042.3765      &   1672    &  10.09    &  0.4680    &   0.0379     &  282.328         &              1      &         0.4261     &      11.7563      &        1       &    12.4266    &  402.3297   &   55042.3784    &     1213    &    9.38     &   0.337    &   0.0319    &   319.457\\
   11956940   &        3    & 55.1881   &  54941.2127     &     241     &  5.01    &  0.7623    &   0.0156   &     56.523          &              1       &        3.2513      &      6.6277       &       0        &      null      &    null       &     null     &    null     &   null     &    null    &     null        &  null\\
   11956947   &        3   & 155.0350   &  54947.7909     &      49     &  7.92     & 0.2537    &   0.0063     &  145.679       &                1        &       9.1021      &      0.0202        &      0        &      null      &    null       &     null      &   null    &    null     &    null     &    null      &    null\\
   11956980   &        3   & 361.2025    & 55238.6510      &     26     &  7.16     & 0.7424    &   0.0050   &    261.221        &               1        &       1.7322      &      0.1056        &      0       &       null     &     null      &      null     &    null    &    null      &   null     &    null    &      null\\
   11957042    &       3   & 362.8912   &  55004.2783     &    2602   &    8.45     & 0.6520     &  0.0491    &   269.575       &                1        &       3.5924      &     11.3753      &        1      &     11.3544   &   362.8965   &   55004.2656     &    1729     &  10.14    &    0.949    &   0.0514   &    123.476\\
   11957046    &       3   & 129.9348   &  54914.3745     &     156     &  6.39     & 0.7150    &   0.0123    &   111.392        &               1        &       0.6795       &     1.4373        &      0      &        null      &    null       &     null    &     null    &    null    &     null    &     null    &      null\\
 ... & ... & ... & ...& ...& ...& ...& ...& ...& ...& ...& ...& ...& ...& ...& ...& ...& ...& ...& ... & ... \\
 ... & ... & ... & ...& ...& ...& ...& ...& ...& ...& ...& ...& ...& ...& ...& ...& ...& ...& ...& ... & ... \\
\hline
\hline
\end{tabular}
\label{tab:results}
\end{sidewaystable}

\section{Results}
\label{sec:results}

Table \ref{tab:results} contains the results of the SOC 9.2 pipeline performance on the suite of injected transit signals. 
A ÔsuccessfulÕ detection is defined as one with a measured orbital period within 3\% of the injected period (in practice, recovered periods are almost entirely within 0.01\% of the injected period), and a measured epoch within 0.5 days of the injected epoch; on inspection these values captured all reasonable matches, see Figure 5 of \citet{Christiansen2015a}. Successful detections are indicated in Table \ref{tab:results} in the RF (recovered flag) column with a value of 1. For these targets, the parameters of the injected transit as recovered by the pipeline are also given, for comparison with the injected parameters. In addition to the successfully recovered injections, 805 targets were identified at an integer alias of the injected period. For the purposes of this experiment they are not defined as ÔsuccessfulÕ detections, but in Table \ref{tab:results} are separately identified in the RF column with a value of 2. In Appendix \ref{app:A} we describe how to generate detection efficiencies such as those described below for a sample of injections from this table, which can be selected across any custom stellar or planetary parameter space. 

\begin{figure}[t!]
\centering
\includegraphics[width=\columnwidth]{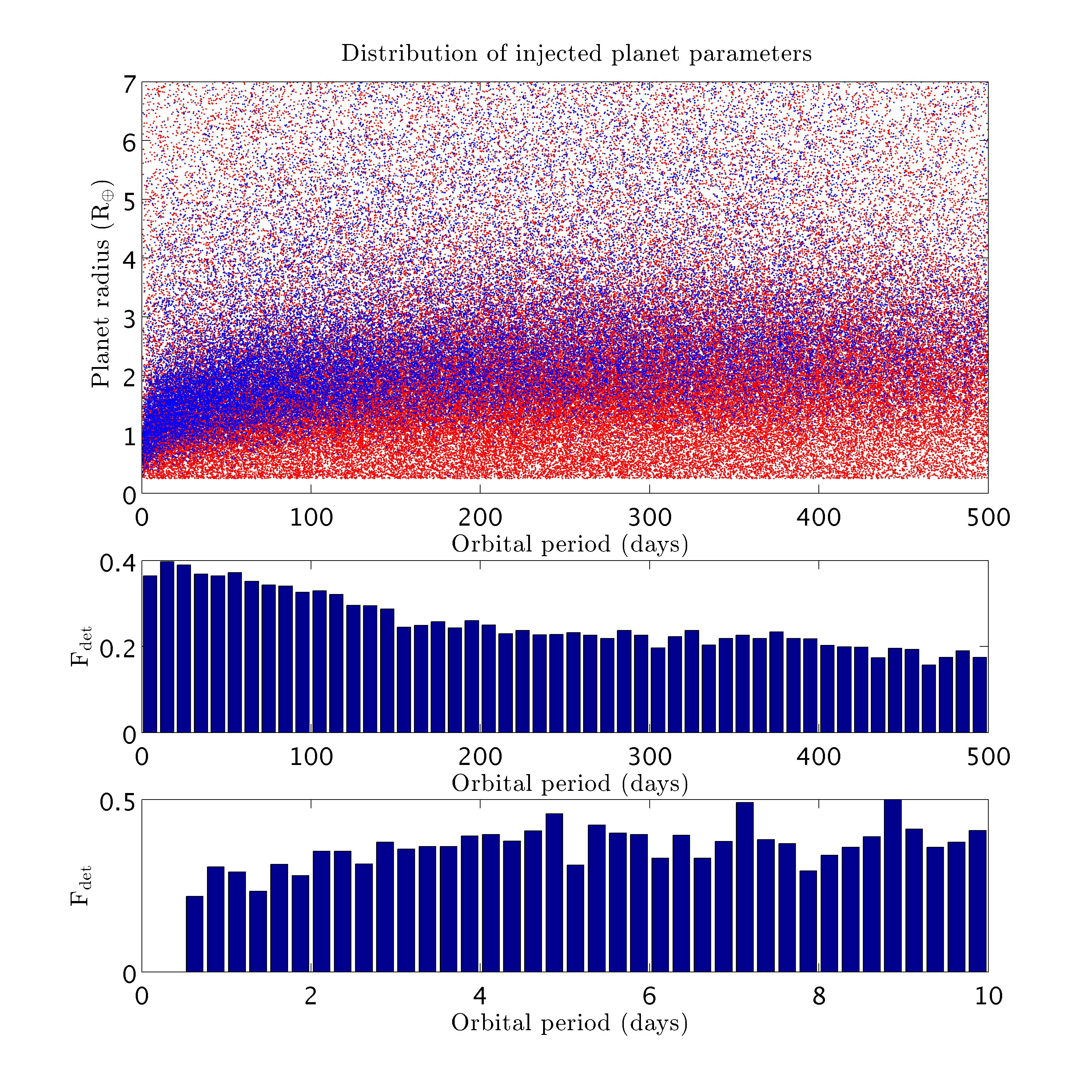}
\caption{The distribution of parameters of the injected and recovered transit signals for all injections. The red points show the signals that were not successfully recovered, and the blue points show the recovered signals.}
\label{fig:injectedparameters}
\end{figure}

The upper panel of Figure \ref{fig:injectedparameters} shows the distribution of injected planet parameters for all 159,013 injections, where the blue points are the injections which are successfully recovered, and the red points are those which are not. The two histograms below show the fraction of injected planets that were successfully recovered as a function of period, over the full 0.5--500 day period range (middle panel) and expanded over the 0.5--10 day period range (lower panel). Note that these histograms include those injections which are not expected to reach the pipeline detection threshold; the median expected detection statistic\footnote{The calculation of the expected detection statistic (MES) includes the following effects:  (i) the noise properties of the flux time series, as described by the Combined Differential Photometric Precision (CDPP; Christiansen et al. 2012\nocite{Christiansen2012}); (ii) the central transit depth; (iii) the dilution of the transit signal by additional flux in the photometric aperture; (iv) the duty cycle of the observations, discarding gapped and deweighted cadences (i.e., those with weights $<0.5$); and (v) the mismatch between the duration of the injected signal and the discrete set of 14 pulse durations searched by the pipeline. Transit signals in the data are compared with test signals of duration 1.5, 2.0, 2.5, 3.0, 3.5, 4.5, 5.0, 6.0, 7.5, 9.0, 10.5, 12.0, 12.5 and 15 hours. Therefore a transit signal with a duration of 3.6 hours, which would have its highest detection statistic when compared to a test signal of duration 3.6 hours, will be measured at 3.5 hours with a slightly lower signal strength.} of the injected planets is 6.5$\sigma$, and the pipeline-based detection threshold is 7.1$\sigma$. We inject many planets both above and below the detection threshold in order to characterise the transition from non-detection to detection. The slight drop in detection efficiency at periods shorter than 4 days seen in the bottom panel of Figure \ref{fig:injectedparameters} is the previously reported effect of the removal of harmonic signatures prior to the periodic signal search \citep{Tenenbaum2012}, which becomes increasingly deleterious of transit signals with shorter periods \citep{Christiansen2013,Christiansen2015a}. The drop in detection efficiency with increasing periods is analysed in more detail below. For the analysis presented below we discard injections that did not result in at least three transits injected on good (not gapped or heavily deweighted) cadences, so the drop in detection efficiency is not a result of the window function of the data (i.e. longer period injections being less likely to result in the required three transits).

For the following analysis, we consider only the simulated transit signals injected at the location of the target star, and restrict those target stars to FGK main sequence stars. Using the Q1--Q17 DR24 \kepler\ stellar properties catalog presented in \citet{Huber2014}, we select targets with stellar effective temperatures between 4000--7000K, and surface gravities greater than 10000cm/s$^2$; this sample comprises 105,184 injections. These are the target stars on which the \kepler\ project is focused for calculating occurrence rates.

\begin{figure}[h!]
\centering
\includegraphics[width=\columnwidth]{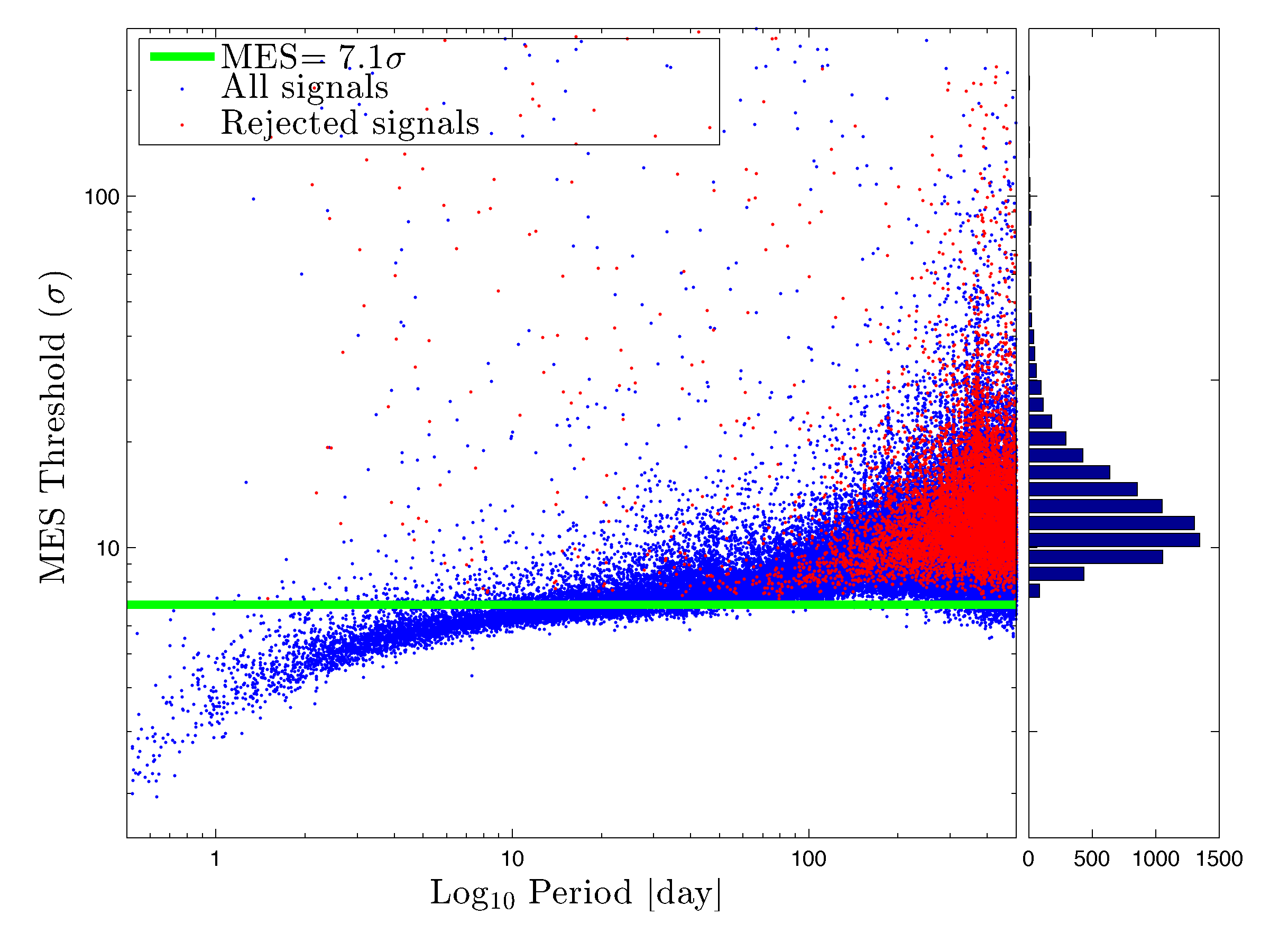}\caption{The statistical bootstrap threshold calculated for each target light curve, for the trial transit duration closest to the duration of the injected transit signal. The bootstrap threshold is a strong function of period. The effective search threshold for each light curve is the larger of the pipeline-based detection threshold (MES$=$7.1$\sigma$, shown as the solid green line) and the statistical bootstrap threshold. The red points designate the signals rejected for having a measured detection statistic below the bootstrap threshold. For periods below 40 days there are few rejections ($\sim$5\% of signals). For periods longer than 40 days the rejection rate rises to $\sim$28\%. The right panel shows a histogram of the statistical bootstrap threshold values for periods longer than 200 days.}
\label{fig:bootstrapmetric}
\end{figure}

In order to generate a detection in TPS, a candidate signal in a target light curve is subjected to four tests. First, the measured MES\footnote{The expected MES of the signal is calculated using the average noise properties of the light curve, however the measured MES is affected by the local noise properties where each transit is injected. On average the measured MES tracks very closely to the expected MES, with a large scatter: 40\% on average for expected MES values of below 20, and 10\% on average for higher expected MES values.} of the signal must be higher than the pipeline-based detection threshold of 7.1$\sigma$. Second, the measured MES must also be higher than the bootstrap detection threshold calculated for that target light curve; this threshold differs from target to target because it depends on the intrinsic noise properties of each light curve. For this particular version of the pipeline, the calculation contained a design flaw and the bootstrap detection thresholds were incorrect, with a bias towards over-estimating the required threshold for low significance signals (typically MES$<$10$\sigma$). As a result the bootstrap test erroneously removed a significant fraction of the candidate signals in this regime that should not have failed, and erroneously passed a somewhat smaller but significant fraction that should not have passed this test. The calculated thresholds also included a large stochastic uncertainty, generating a much wider distribution of thresholds than expected; Figure \ref{fig:bootstrapmetric} shows that some thresholds were erroneously set higher than 100$\sigma$. As a result, we cannot apply a systematic correction to the bias such that we might reproduce the previous behaviour of the pipeline. Finally, if the measured MES is higher than both the bootstrap and pipeline-based detection thresholds, the signal is tested against the remaining vetoes: the robust statistic veto (Tenenbaum et al. 2013); and the $\chi^2_2$ and $\chi^2_{GOF}$ vetoes (Seader et al. 2013, 2015a). The former compares individual transit events to the phased transit signal folded at the trial period and penalises those which differ significantly, in order to remove cases where a large outlier or systematic artefact in the light curve is folded onto two much shallower events and generates a significant detection above the previous thresholds. The $\chi^2$ vetoes compare the individual transit events to a physical transit template and penalise those which are not a good match. If it passes these vetoes, it is considered by the pipeline to be a detection; a successful detection is one which also matches the period and epoch of the injected signal as defined above. 


Figure \ref{fig:bootstrapmetric} shows a comparison of the bootstrap and pipeline-based detection thresholds for each injection. For periods shorter than 40 days, the bootstrap detection threshold is typically below the pipeline-based detection threshold (the solid green line) and therefore the vast majority (95\%) of the light curves are searched down the MES$=$7.1$\sigma$ threshold, and then tested against the additional vetoes. The detection efficiency in this period range therefore behaves as previously, in that the pipeline sensitivity can be described by a uniform search down to a given detection threshold.

\begin{figure}[h!]
\centering
\includegraphics[width=\columnwidth]{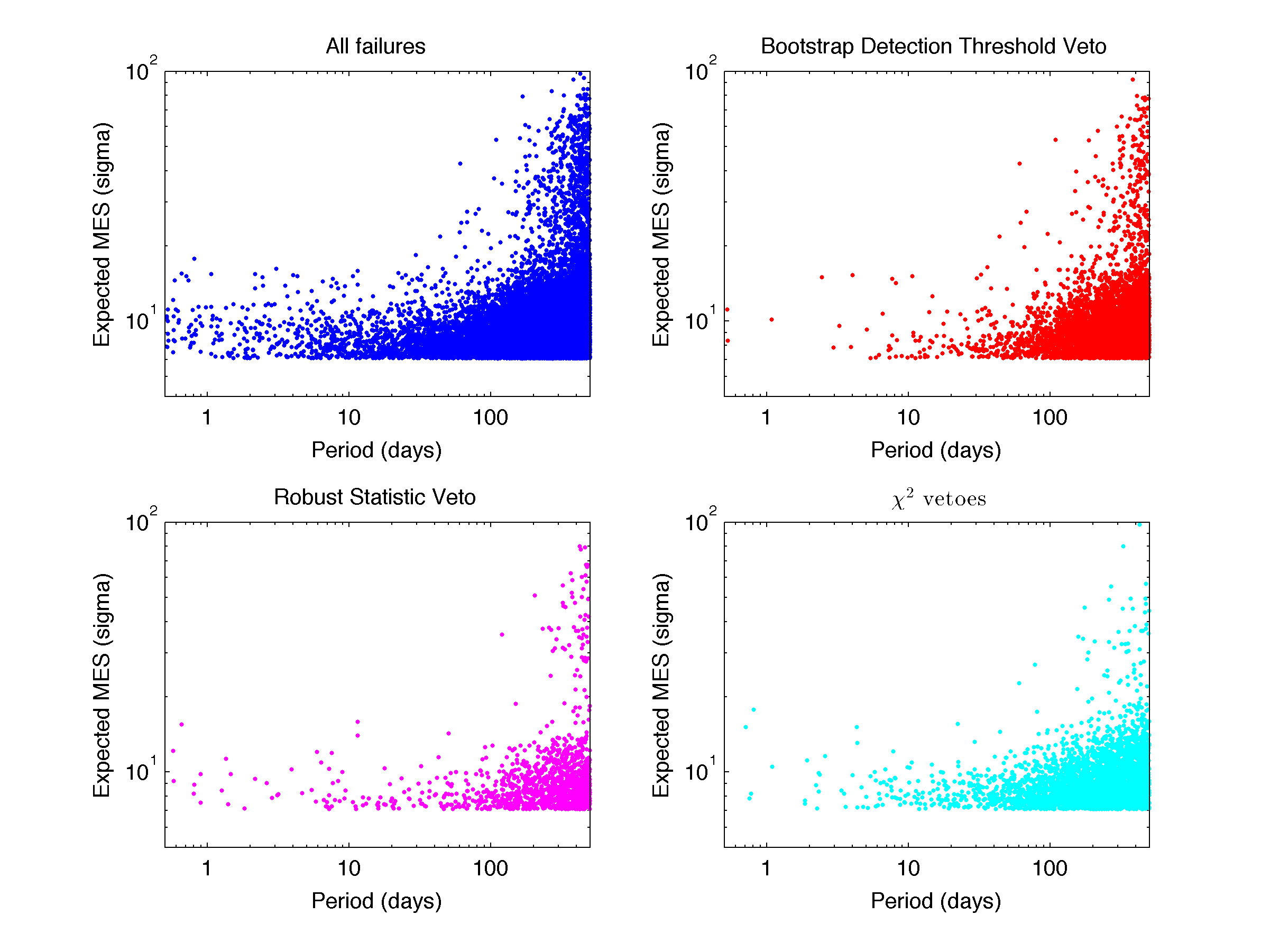}\caption{The distribution of the parameters of the injected signals removed by the vetoes. The top left panel shows all non-detected signals with expected detection statistics above the pipeline-based detection threshold of MES$=$7.1$\sigma$. This includes signals with measured detection statistics below the pipeline-based detection threshold which were subsequently not subjected to the vetoes. The top right panel shows the parameters of the injected signals removed by the bootstrap veto (5597 in total). The bottom left panel shows which signals were removed by the next veto to act, the robust statistic veto (1247 in total). The bottom right panel shows which signals were removed by the final two vetoes, the $\chi^2_2$ and $\chi^2_{GOF}$ vetoes (3647 in total).}
\label{fig:vetoes}
\end{figure}

For orbital periods longer than 40 days, the bootstrap detection threshold increases above the pipeline-based detection threshold to a median of $\sim$11 and shows large scatter from target to target. As a result, the bootstrap veto rejects a large number of the injected signals, rising from $\sim$5\% with periods less than 40 days to $\sim$28\% for periods longer than 300 days; in total, 5597 signals are removed, 5483 of those with periods above 40 days. The majority of the rejected signals (93\%) have expected detection statistics below 15$\sigma$ (99\% have measured detection statistics below 15$\sigma$). The large scatter of the bootstrap detection thresholds from target to target and strong period dependence for the resulting rejections violate the assumptions of \citet{Burke2015}, which precludes the use of a derived average detection efficiency as justified previously. The other vetoes in TPS subsequently remove an additional $\sim$4600 signals at periods longer than 40 days, however we cannot usefully characterise their sensitivity due to the prior rejection of a large number of signals by the bootstrap veto. The distribution of the injected signals removed by each of the vetoes in turn is illustrated in Figure \ref{fig:vetoes}.

\begin{figure}[h!]
\centering
\includegraphics[width=\columnwidth]{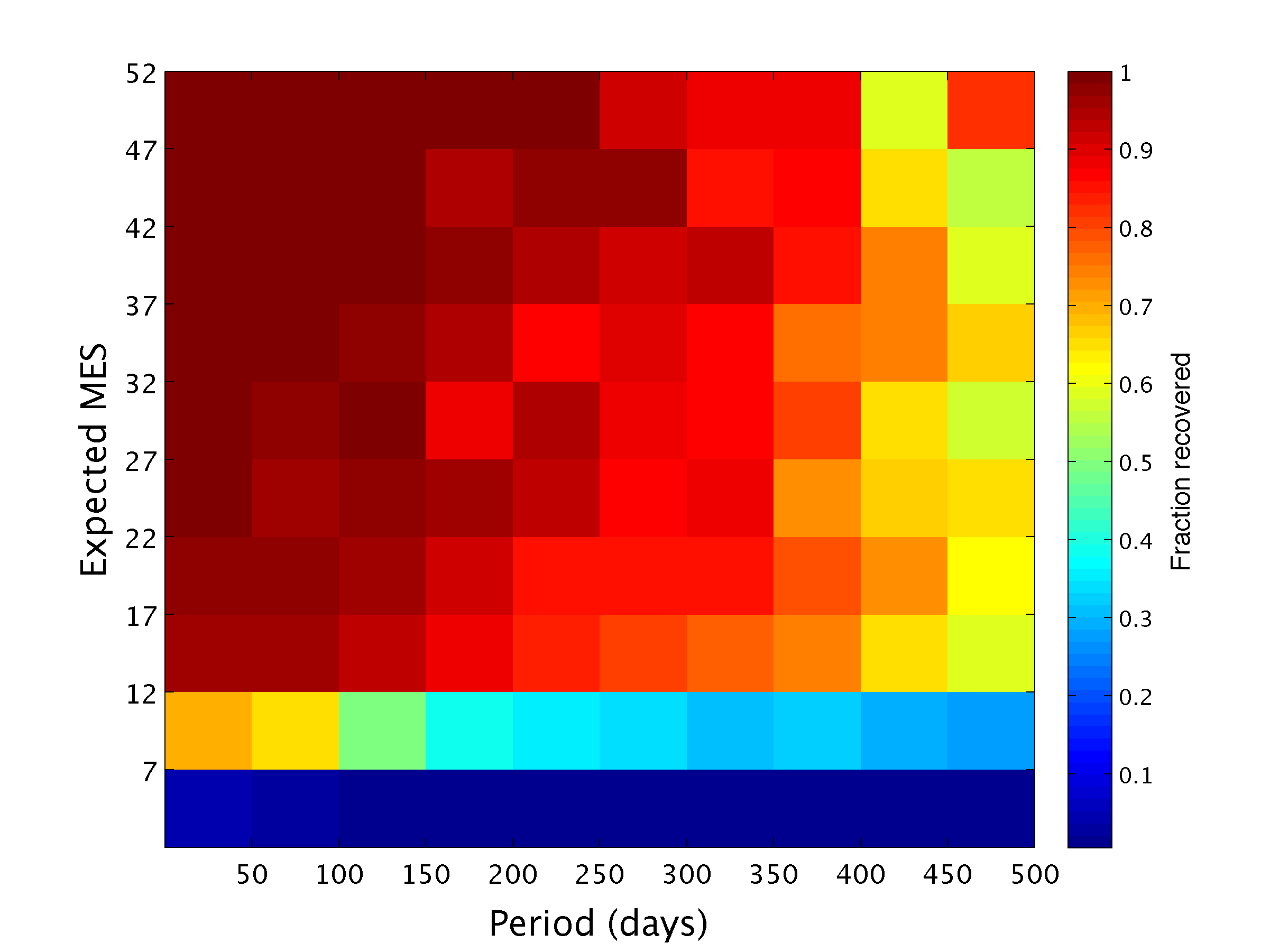}\caption{The fraction of injected signals successfully recovered by the pipeline, for the FGK dwarfs (4000K $< T_{\rm eff} <$ 7000K, log $g>4.0$; 105,184 injections in total). Note the marked drop-off in detectability below the pipeline-based detection threshold of MES$=$7.1$\sigma$. For periods longer than 150 days, the sensitivity falls off even at high MES values.}
\label{fig:twodsensitivity}
\end{figure}

Figure \ref{fig:twodsensitivity} shows the resulting two-dimensional dependence of the pipeline sensitivity, as a function of both the expected detection statistic, and the orbital period of the injection, for all 105,184 injections considered. We see the marked decrease in the pipeline sensitivity at longer orbital period, falling from $\sim$90\% completeness at periods shorter than 150 days to below 70\% at periods longer than 400 days. 

\begin{figure}[h!]
\centering
\includegraphics[width=\columnwidth]{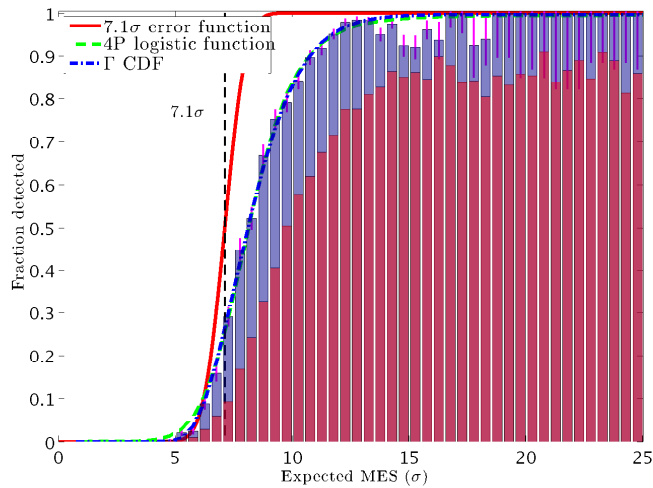}\caption{The detection efficiency of the \kepler\ SOC 9.2 pipeline as a function of the expected detection statistic of the injected transit signal (expected MES) using the Q1--Q17 DR 24 light curves. The blue histogram shows the efficiency for periods less than 40 days, and the red for periods longer than 40 days. The black dashed line shows the pipeline-based detection threshold of MES$=$7.1$\sigma$. The solid red line is the hypothetical performance of the detector on perfectly whitened noise, which is an error function centred on MES$=$7.1$\sigma$. The dot-dashed blue line is the gamma cumulative distribution function fit to the histogram, and the dashed green line is the four-parameter logistic fit to the histogram. The magenta bars show the uncertainty in each bin assuming a binomial distribution.}
\label{fig:sensitivitycurve}
\end{figure}

Following from Figures \ref{fig:bootstrapmetric} and \ref{fig:twodsensitivity}, the prescription outlined in previous work for characterising the detection efficiency of the pipeline simply as a function of the expected detection statistic is invalid for periods $>$40 days. For the injections with periods shorter than 40 days, where we do not expect the statistical bootstrap metric to affect the detection efficiency, we derive the detection efficiency in a similar fashion to that described in \citet{Christiansen2015a}, shown in Figure \ref{fig:sensitivitycurve}. As previously, we fit a $\Gamma$ cumulative distribution function of the form

\begin{equation}
	p = F(x|a,b)=\frac{c}{b^a\Gamma(a)}\int\limits_0^x t^{a-1}e^{-t/b}dt
\end{equation}

\noindent where $p$ is the probability of detection, $\Gamma$(a) is the gamma function, $x = MES$, and $c$ is a scaling factor such that the maximum detection efficiency is the average of the per-bin detection probabilities recovered for $15 < MES < 50.$ The use of the gamma function is common in describing the rate of physical processes, in this case the detection of the injected signal. A fit of this function to the histogram, shown in Figure \ref{fig:sensitivitycurve} as the solid green line, gives coefficients $a = 23.11$, $b = 0.36$, and $c = 0.997$. For comparison we also fit a four-parameter logistic function of the form $F(x|a_l,b_l,c_l,d_l) = ((a_l - d_l)/(1+(x/c_l)^{b_l})+ d_l$, where we fix $a_l$ (the minimum sensitivity) to 0, and a fit to the histogram, shown as the solid cyan line, gives $b_l=8.06$, $c_l=8.11$, and $d_l=0.995$. The two fits (both with three free parameters) give very similar reduced $\chi^2$ values (1.00 and 1.07) respectively, where the uncertainties in each histogram bin are calculated assuming a binomial distribution. For these short period injections, the recovery rate of strong signals, with expected detection statistics $>15$, is very close to unity ($>99.5$\%) as expected. 

One area of investigation is whether the presence of multiple planetary signals in a given light curve affects the detection efficiency of each individual signal. This could occur if, for instance, the presence of many transit signals increased the noise properties of the light curve such that individual signals were detected with lower significance. In addition, the order in which  signals are detected will influence their detectability: since candidate transit signals are removed after they are detected and before the light curve is searched again, shorter period signals typically remove more observations than longer period signals, affecting the window function of subsequent searches.

The simplest check is to remove the 3357 targets with planet candidates identified with the 9.2 pipeline from the 105,184 targets and repeat the above calculation. This is a relatively small number to remove, and the derived parameters are effectively unchanged for periods shorter than 40 days. The new $\Gamma$ function coefficients are $a = 23.26$, $b = 0.36$, and $c = 0.996$, and the new logistic function coefficients are $b_l=8.08$, $c_l=8.11$, and $d_l=0.994$. There are too few injections with periods below 40 days around known planet candidate hosts (176 in total) to examine the detection efficiency of signals in light curves with known additional signals, and we defer that analysis, and a more extensive examination of this effect in general for the full, robust data set.

\section{Conclusions}
\label{sec:conclusions}

Previously, we had generated a simple prescription to describe the detection efficiency of the pipeline as a function of the expected detection statistic, subsequently used in \citet{Burke2015} and \citet{Christiansen2015a} to robustly calculate planet occurrence rates. Due to the statistical bootstrap metric introduced in SOC version 9.2 of the \kepler\ pipeline, we are unable to regenerate this prescription except for periods shorter than 40 days. As was demonstrated in those previous papers, incorrect assertions about the detection efficiency can introduce very large systematic errors in the derived occurrence rates. We therefore recommend strongly that the Q1--Q17 DR24 planet candidate catalogue presented in \citet{Coughlin2015}, which was produced with SOC version 9.2, be used to calculate occurrence rates only for orbital periods shorter than 40 days.

The adverse behaviour described here is isolated to SOC version 9.2 and does not impact the previous SOC version 9.1 results, including the Q1--Q16 planet candidate catalogue presented in \citet{Mullally2015}; see Section 7 of that paper for relevant caveats as to the completeness and reliability of that catalogue. The design flaw in the SOC version 9.2 bootstrap code has been identified and corrected \citep{Jenkins2015}. The corrected statistical bootstrap metric and associated values have been archived at the NASA Exoplanet Archive with the Q1--Q17 DR24 TCE catalog and are documented in \citet{Seader2015b}. Additionally, the SOC 9.3 transit search code (TPS) has been further modified to reduce other sources of bias \citep{Jenkins2015}. This includes changing the use of the statistical bootstrap metric from a veto (rejecting signals from further consideration) to a vetting diagnostic (used in classifying events into likely planet candidates or false positives after the events have been identified by the \kepler\ pipeline). Therefore, the SOC version 9.3 DR25 KOI catalog should be amenable to occurrence rate calculations using the prescription in \citet{Burke2015} and \citet{Christiansen2015a}. 

\acknowledgments

Funding for the \kepler\ Discovery Mission is provided by NASA's Science Mission Directorate. The authors acknowledge the efforts of the Kepler Mission team for obtaining the calibrated pixels, light curves and data validation diagnostics data used in this publication. These data products were generated by the Kepler Mission science pipeline through the efforts of the Kepler Science Operations Center and Science Office. The Kepler Mission is lead by the project office at NASA Ames Research Center. Ball Aerospace built the Kepler photometer and spacecraft which is operated by the mission operations center at LASP. These data products are archived at the Mikulski Archive for Space Telescopes and the NASA Exoplanet Archive. JLC is supported by NASA under award No. GRNASM99G000001.

\appendix
\section{A Suggested Recipe for Calculating the Average Pipeline Detection Efficiency}
\label{app:A}

Here we outline one process for determining the pipeline detection efficiency as a function of the expected detection statistic (MES), using the full table of injections and recoveries described in the text and available at the NASA Exoplanet Archive. This allows the reader to calculate the likelihood that the pipeline would have detected a transit at a given signal to noise. If one is interested in particular regions of planet and stellar parameter space, one can then calculate the signal to noise of the candidate signals and compute their recovery rates. 

\begin{enumerate}
\item Select a detection threshold above which to calculate the detection efficiency. The default is the standard pipeline-based detection threshold (MES$=$7.1$\sigma$; Jenkins 2002) and this represents the minimum threshold valid for this procedure. For periods longer than 40 days, we recommend selecting a higher (MES=15-20) threshold. If a new, higher threshold is chosen, change the ÔrecoveredÕ flag (column 13 of the results table) to 0 for objects from the table with measured MES (column 14) below the threshold, recognising that they would not have been detected under the higher threshold. Otherwise keep all rows to reproduce the standard MES$=$7.1$\sigma$ threshold.
\item Select the parameter space in stellar and/or planet properties over which to calculate the detection efficiency; for the analysis described here, we selected FGK main sequence stars. The Kepler stellar properties table available at the NASA Exoplanet Archive can be used to identify which Kepler IDs (column 1 of the results table) fall into a given stellar parameter range. To select over desired planet properties, use columns 3--9 in the table to remove injections that fall outside the desired parameter space.
\item Finally, for occurrence rate calculations, choose the subset of targets that were injected on-target using the flag in column 10 of the results table (simulating transiting planets on the target star). For certain false positive rate investigations (e.g., Mullally et al. 2015b), instead use those targets that were injected at a location offset from the target star.
\item Select your desired expected MES (column 12 in the results table) bins (for the analysis in Figure \ref{fig:sensitivitycurve} we examine MES from 0-100 with bins of width 0.5). For each bin, $i$, count the number of targets in the final set of rows from the now truncated table with an expected MES falling in that bin, $N_{i,\rm{exp}}$, and of those, the number that were successfully recovered, $N_{i,\rm{det}}$, using either the flag in column 13 if you are using the standard MES$=$7.1$\sigma$ threshold, or by imposing the condition that the measured MES (column 14) be greater than your chosen threshold. Then calculate the detection efficiency $N_{i,\rm{det}}/N_{i,\rm{exp}}$ for each bin. 
\item Calculate a histogram of the resulting detection efficiency and fit a function of your choice to the histogram values. We have found both cumulative $\Gamma$ distribution functions and four-parameter logistic functions to fit well.
\item Use the function to correct the completeness rates in your occurrence rate calculation; see the text for strong caveats on where and how this is a valid correction for SOC version 9.2.
\end{enumerate}




\clearpage

\begin{thebibliography}{}

\bibitem[Akeson et al.(2013)]{Akeson2013} Akeson, R.~L., Chen, X., Ciardi, D., et al.\ 2013, \pasp, 125, 989 
\bibitem[Bryson et al.(2013)]{Bryson2013} Bryson, S.~T., Jenkins, J.~M., Gilliland, R.~L., et al.\ 2013, \pasp, 125, 889 
\bibitem[Burke et al.(2015)]{Burke2015} Burke, C.~J., Christiansen, J.~L., Mullally, F., et al.\ 2015, \apj, 809, 8 
\bibitem[Christiansen et al.(2012)]{Christiansen2012} Christiansen, J.~L., Jenkins, J.~M., Caldwell, D.~A., et al.\ 2012, \pasp, 124, 1279
\bibitem[Christiansen et al.(2013)]{Christiansen2013} Christiansen, J.~L., Clarke, B.~D., Burke, C.~J., et al.\ 2013, \apjs, 207, 35
\bibitem[Christiansen et al.(2015a)]{Christiansen2015a} Christiansen, J.~L., Clarke, B.~D., Burke, C.~J., et al.\ 2015a, \apj, 810, 95 
\bibitem[Christiansen et al.(2015b)]{Christiansen2015b} Christiansen, J.~L. 2015b, ``KSCI-19094-001: Planet Detection Metrics: Pipeline Detection Efficiency'', {\url{http://exoplanetarchive.ipac.caltech.edu/docs/KSCI-19094-001.pdf}}
\bibitem[Coughlin et al.(2016)]{Coughlin2015} Coughlin, J.~L., Mullally, F., Thompson, S.~E., et al.\ 2016, \apj, accepted arXiv:1512.06149 
\bibitem[Coughlin et al.(2014)]{Coughlin2014} Coughlin, J.~L., Thompson, S.~E., Bryson, S.~T., et al.\ 2014, \aj, 147, 119 
\bibitem[Dressing \& Charbonneau(2015)]{Dressing2015} Dressing, C.~D., \& Charbonneau, D.\ 2015, \apj, 807, 45 
\bibitem[Huber et al.(2014)]{Huber2014} Huber, D. 2014, ``KSCI-19083-001: Kepler Stellar Properties Catalog Update
for Q1-Q17 Transit Search'', {\url{http://exoplanetarchive.ipac.caltech.edu/docs/KeplerStellar_Q1_17_documentation.pdf}}\bibitem[Jenkins(2002)]{Jenkins2002} Jenkins, J.~M.\ 2002, \apj, 575, 493 
\bibitem[Jenkins et al.(2010a)]{Jenkins2010a} Jenkins, J.~M., Caldwell, D.~A., Chandrasekaran, H., et al.\ 2010a, \apjl, 713, L87 
\bibitem[Jenkins et al.(2010b)]{Jenkins2010b} Jenkins, J.~M., Chandrasekaran, H., McCauliff, S.~D., et al.\ 2010b, \procspie, 7740,  10
\bibitem[Jenkins et al.(2015)]{Jenkins2015} Jenkins, J.~M., Twicken, J.~D., Batalha, N.~M., et al.\ 2015, \aj, 150, 56
\bibitem[Kane et al.(2014)]{Kane2014} Kane, S.~R., Kopparapu, R.~K., \& Domagal-Goldman, S.~D.\ 2014, \apjl, 794, L5
\bibitem[Mandel \& Agol(2002)]{Mandel02} Mandel, K. \& Agol, E.\ 2002, \apj, 580, 171
\bibitem[Mullally et al.(2015)]{Mullally2015} Mullally, F., Coughlin, J.~L., Thompson, S.~E., et al. 2015, \apjs, 217, 31  
\bibitem[Mullally et al.(2016)]{Mullally2016} Mullally, F., Coughlin, J.~L., Thompson, S.~E., et al.\ 2016, \pasp, accepted arXiv:1602.03204 
\bibitem[Quintana et al.(2010)]{Quintana2010} Quintana, E.~V., Jenkins, J.~M., Clarke, B.~D., et al.\ 2010, \procspie, 7740, 64
\bibitem[Santerne et al.(2015)]{Santerne2015} Santerne, A., Moutou, C., Tsantaki, M., et al.\ 2015, arXiv:1511.00643 
\bibitem[Seader et al.(2013)]{Seader2013} Seader, S., Tenenbaum, P., Jenkins, J.~M., \& Burke, C.~J.\ 2013, \apjs, 206, 25 
\bibitem[Seader et al.(2015a)]{Seader2015a} Seader, S., Jenkins, J.~M., Tenenbaum, P., et al.\ 2015, \apjs, 217, 18 
\bibitem[Seader et al.(2015b)]{Seader2015b} Seader, S., Jenkins, J.~M.,  \& Burke, C. 2015, Planet Detection Metrics: Statistical Bootstrap Test (KSCI-19086) 
\bibitem[Smith et al.(2012)]{Smith2012} Smith, J.~C., Stumpe, M.~C., Van Cleve, J.~E., et al.\ 2012, \pasp, 124, 1000 
\bibitem[Stumpe et al.(2012)]{Stumpe2012} Stumpe, M.~C., Smith, J.~C., Van Cleve, J.~E., et al.\ 2012, \pasp, 124, 985 
\bibitem[Stumpe et al.(2014)]{Stumpe2014} Stumpe, M.~C., Smith, J.~C., Catanzarite, J.~H., Van Cleve, J.~E., Jenkins, J.~M., Twicken, J.~D., \& Girouard, F.~R.\ 2014, \pasp, 126, 100
\bibitem[Tenenbaum et al.(2012)]{Tenenbaum2012} Tenenbaum, P., Christiansen, J.~L., Jenkins, J.~M., et al.\ 2012a, \apjs, 199, 24 
\bibitem[Tenenbaum et al.(2013)]{Tenenbaum2013} Tenenbaum, P., Jenkins, J.~M., Seader, S., et al.\ 2013, \apjs, 206, 5 
\bibitem[Tenenbaum et al.(2014)]{Tenenbaum2014} Tenenbaum, P., Jenkins, J.~M., Seader, S., et al.\ 2014, \apjs, 211, 6 
\bibitem[Thompson et al.(2015)]{Thompson2015} Thompson, S.~E., Mullally, F., Coughlin, J., et al.\ 2015, \apj, 812, 46 
\bibitem[Twicken et al.(2010)]{Twicken2010} Twicken, J.~D., Clarke, B.~D., Bryson, S.~T., et al.\ 2010, \procspie, 7740,  774023
\bibitem[Wu et al.(2010)]{Wu2010} Wu, H., Twicken, J.~D., Tenenbaum, P., et al.\ 2010, \procspie, 7740, 42

\end{thebibliography}
\end{document}